\documentclass{bmcart}

\usepackage{amsthm,amsmath,amssymb}
\usepackage{graphicx}
\usepackage[utf8]{inputenc}

\begin{document}

\begin{frontmatter}

\begin{fmbox}
\dochead{Research}

\title{Emergence of spatial transitions in urban congestion dynamics}

\author[
  addressref={aff1,aff2},
  corref={aff1,aff2},
  email={nello.lampo@gmail.com}
]{\inits{A.L.}\fnm{Aniello} \snm{Lampo}}
\author[
  addressref={aff1},
  email={jborgeh@uoc.edu}
]{\inits{J.B.-H.}\fnm{Javier} \snm{Borge-Holthoefer}}
\author[
  addressref={aff2},
  email={sergio.gomez@urv.cat}
]{\inits{S.G.}\fnm{Sergio} \snm{G{\'o}mez}}
\author[
  addressref={aff1,aff3},
  email={asolerib@uoc.edu}
]{\inits{A.S.-R.}\fnm{Albert} \snm{Sol{\'e}-Ribalta}}

\address[id=aff1]{%
  \orgdiv{Internet Interdisciplinary Institute (IN3)},
  \orgname{Universitat Oberta de Catalunya},
  \city{Barcelona},
  \cny{Catalonia, Spain}
}
\address[id=aff2]{%
  \orgdiv{Departament d'Enginyeria Inform\`{a}tica i Matem\`{a}tiques},
  \orgname{Universitat Rovira i Virgili},
  \city{Tarragona},
  \cny{Catalonia, Spain}
}
\address[id=aff3]{%
  \orgdiv{URPP Social Networks},
  \orgname{University of Zurich},
  \city{Zurich},
  \cny{Switzerland}
}

\end{fmbox}

\begin{abstractbox}
\begin{abstract}
The quantitative study of traffic dynamics is crucial to ensure the efficiency of urban transportation networks. The current work investigates the spatial properties of congestion, that is, we aim to characterize the city areas where traffic bottlenecks occur. The analysis of a large amount of real road networks in previous works showed that congestion points experience spatial abrupt transitions, namely they shift away from the city center as larger urban areas are incorporated. The fundamental ingredient behind this effect is the entanglement of central and arterial roads, embedded in separated geographical regions.

In this paper we extend the analysis of the conditions yielding abrupt transitions of congestion location. First, we look into the more realistic situation in which arterial and central roads, rather than lying on sharply separated regions, present spatial overlap. It results that this affects the position of bottlenecks and introduces new possible congestion areas. Secondly, we pay particular attention to the role played by the edge distribution, proving that it allows to smooth the transitions profile, and so to control the congestion displacement. Finally, we show that the aforementioned phenomenology may be recovered also as a consequence of a discontinuity in the node's density, in a domain with uniform connectivity. Our results provide useful insights for the design and optimization of urban road networks, and the management of the daily traffic.
\end{abstract}


\end{abstractbox}

\end{frontmatter}

\section{Introduction}
\label{Intro}

The sustainability of urban life represents one of the greatest challenges of our time. This has sparked a lot of interest among researchers coming from several areas, and has led to the birth of the so-called science of cities \cite{Batty2012}. In this context, the engineering and optimization of road transportation networks stands as a crucial task to ensure the efficiency of traffic dynamics and to control the related congestion emergence.

In the literature, the study of road transportation networks has been addressed following different methods, according to the purpose and the system characteristics. For instance, the phenomenology emerging on arterial roads (also named inter-urban, peripheral or high-capacity roads), characterized by long segments and limited inter-connections, has been treated by means of fluid models \cite{helbing2015traffic}, or the fundamental diagram of traffic flow \cite{godfrey1969mechanism, daganzo2008analytical}. The case of central streets (also called intra-urban), instead, has been traditionally analyzed by means of network science \cite{Guimera2002,Sole2018,chen2020traffic}.

These two types of road networks, which are usually studied independently, are increasingly intertwined as cities sprawl over suburban areas. Therefore, an entanglement perspective is of great importance to properly deal with the dynamics of vehicles flux, as well as other related phenomena. In \cite{Lampo2021}, we pursued this path and focused on the spatial behavior of congestion onset. We found that the structural discontinuity between the central and arterial roads yields an abrupt transition of the location where congestion occurs. In particular, congestion points shift away from the city center as larger areas are included in the urban spatial domain. Remarkably, such property is also detected in almost one hundred real cities road networks worldwide.

This result represents an important resource to improve the efficiency of road networks. The displacement of the traffic bottlenecks towards the peripheral zones, indeed, is crucial to reduce the pressure on the city center, and to avoid the degradation of the transportation system. Therefore, the possibility to control and manipulate such effect constitutes a notable task which has never been performed before. The current paper goes in this direction. Specifically, we present a series of extensions of the theoretical setup in \cite{Lampo2021}, aimed to approach more realistic situations, and shed light on the fundamental ingredients ruling the observed phenomenon.

In \cite{Lampo2021}, the spatial separation between urban center and the surrounding periphery was recognized to be a necessary requirement for the emergence of congestion transitions. There, the road transportation networks are treated as the inextricable blending of two different structures, that are however sharply separated in space. Here, instead, we look into the consequences of a spatial overlap between center and periphery, namely we introduce a hybrid region where the dense and highly regular frame of central roads coexists with that of the arterial ones. This is a crucial point to match real contexts, since the transition between center and periphery may not be sharp, but it is expected to occur in a wider spatial range.

Along this line, we investigate the role of the edge distribution, implementing the most general case of a non-uniform spatial-dependent density. This turns to be a key object to manipulate the location of congestion onset, allowing to smooth its shift from the center. The edge distribution, i.e., the number of edges over the unit area, may also be manipulated by modifying the number of nodes. Thus, we pay great attention to the situation in which connectivity is assumed to be uniform, while urban center and periphery are constructed as regions with a high and low concentration of nodes, respectively. In this way, we recover the abrupt transitions of congestion, and highlight the importance of both node and edge distributions to the control of traffic bottlenecks.

The manuscript is organized as follows. In Sec.~\ref{sec:DT_MST_model} we present the DT-MST random planar model developed in \cite{Lampo2021}, constituting a practical benchmark to deal with the interplay of central and arterial roads. The former are described using a Delaunay triangulation (DT), while the latter are treated using a minimum spanning tree (MST), both embedded on separated spatial domains. In Sec.~\ref{sec:transitions} we employ this model to study the behavior of the congestion radius as a function of the urban size, showing that it is possible to detect two regimes, which correspond to bottlenecks located in the center, and in its connection with the periphery, separated by an abrupt transition. In Sec.~\ref{sec:Spatial_Overlap} we consider the more realistic situation where MST and a DT are not sharply separated and can share a finite portion of space. We find that this yields a second transition, as a consequence of the further structural discontinuity introduced by the new hybrid region. In Sec.~\ref{sec:Edge_Dist} we extend the previous cases by including the possibility of a non-uniform edge distribution, and show that it leads to smoother transitions. Finally, in Sec.~\ref{sec:NodeDens} we investigate the inverse situation in which center and periphery emerge as areas with different node densities, but maintain the same connectivity. This situation enables the recovery of the abrupt transition and highlight the role of nodes distribution. The importance of these results for urban planning, as well as possible future extensions, are discussed in Sec.~\ref{sec:Conclusions}.

\section{A model for center-periphery entanglement}
\label{sec:DT_MST_model}

\begin{figure}[tb!]
  \includegraphics[width=0.95\columnwidth]{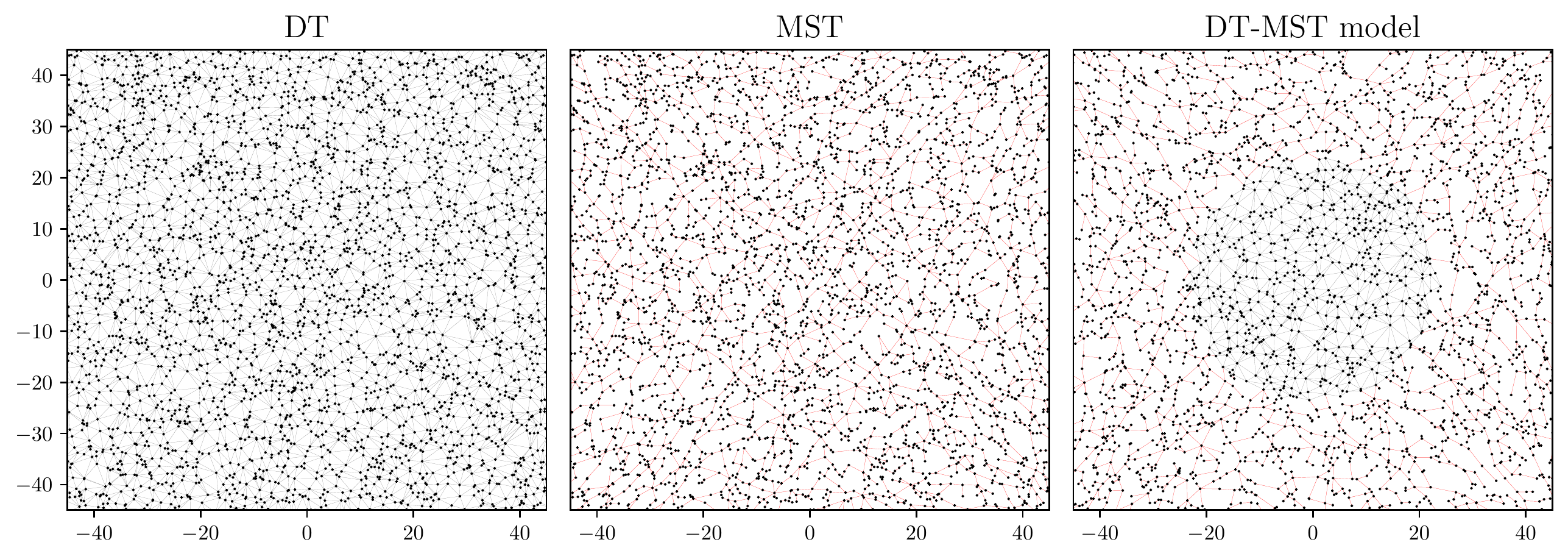}
  \caption{\label{fig:Net_Fig} Graphical representation of a DT (left), MST (middle), and a DT-MST random planar model (right). Networks are built over a uniform distribution of $N=4000$ in a square with $L=45$ side. The DT-MST random planar model is obtained by keeping the DT in a circle with radius $R^{DT}=24$, and the MST outside.}
\end{figure}

Road transportation networks consist of networks whose edges represent roads, while the points where they intersect constitute the nodes. Road networks are a paradigmatic example of planar graphs, namely graphs which can be embedded on a surface in such a way that no edges cross each other; of course, real road networks may have crossing edges, but they always represent a very small fraction of the whole system. In this section we present a general class of planar graphs that, while being as free as possible of any microscopic constraints, keeps the distinctive feature of an entanglement between central and arterial roads.

The most natural choice to model urban center, characterized by a dense web of streets resembling short segments, is provided by a Delaunay triangulation (DT). For a given set of points in a plane, DT is a triangulation over the plane, i.e., any triad of neighboring points defines a triangle, such that no point lies within the circumcircle of any triangle. An example of DT is depicted in Fig.~\ref{fig:Net_Fig} for a random uniform distribution of points. Roughly speaking, DT yields a network structure characterized by a high number of shorts edges, uniformly spread over all the nodes (regularity), with a great redundancy of paths. Remarkably, there are no edge overlapping, i.e., DT preserves planarity. Similar properties could be recovered by considering other types of regular graphs, such as grids, but these would impose a spatial constraint on the nodes positions, making the model contrived. However, as widely discussed in \cite{Lampo2021}, this kind of structure produces equivalents results for congestions.

Peripheral arterial roads are high-capacity streets whose primary function is to deliver traffic between different urban centers. As such, intersections are often reduced to improve traffic flow, and the resulting peripheral street frame resemble a sparse web of long segments. Still, we seek for a network structure where these properties emerge automatically.  The low connectivity suggests to look into a tree structure, whose general implementation is provided by a minimum spanning tree (MST).

The MST is defined as the subset of the edges (of a connected, edge-weighted undirected graph) connecting the vertices in such a way that there are no redundant paths, and minimizing the total edge weight. In our model, the construction of the MST is obtained from a DT, in which the edge weights are the reciprocals of their corresponding edge-betweenness, rather than the Euclidean distance between their adjacent nodes. This is a crucial aspect to reproduce the main routing backbones of arterial periphery webs, as shown in Fig.~\ref{fig:Net_Fig} (middle). A distance-based MST, like that employed in \cite{Kirkley2018}, does not consider routing flows, which are essential in transportation networks.

The model we propose to treat central and arterial peripheral roads in a unified entangled framework is constituted by a DT core surrounded by a MST. Summarizing, the resulting synthetic network is built according to the following steps:
\begin{itemize}
\item We consider a uniform random distribution of $N$ points (the nodes) in the spatial domain $\mathcal{I}=\left[L,-L\right]\times\left[L,-L\right]$;
\item This set of points induces a DT over $\mathcal{I}$. We hold it within the circle of radius $R^{(DT)}$, i.e., for points with radius $r\leq R^{(DT)}$;
\item Out of such a circle, for $r> R^{(DT)}$, we look for the MST resulting from the DT, obtained by minimizing with respect to the reciprocals of the edge betweenness of the links.
\end{itemize}
Our model, coined as DT-MST random planar model, is so defined by only three parameters: the square spatial side length $L$; the number of nodes $N$; and the DT radius, $R^{DT}$. In Fig.~\ref{fig:Net_Fig} (right) we present a realization of the model for a given choice of these parameters.

The idea to model central and peripheral roads as a DT and a MST, respectively, may be better understood in terms of betweenness distribution. This has been carefully studied in \cite{Kirkley2018} for a significant number of cities worldwide, proving that it exhibits an invariant bimodal structure. Such bimodal character is a first indicator of the existence of a structural discontinuity between different topologies as we depart from the city center. Specifically, as discussed in Sec.~III of \cite{Lampo2021}, the two branches of the betweenness distribution may be fitted with those of a tree and a DT respectively, providing an endorsement of the DT-MST model.

\section{Spatial abrupt transitions of congestion}
\label{sec:transitions}

\begin{figure}[tb!]
  \includegraphics[width=0.95\columnwidth]{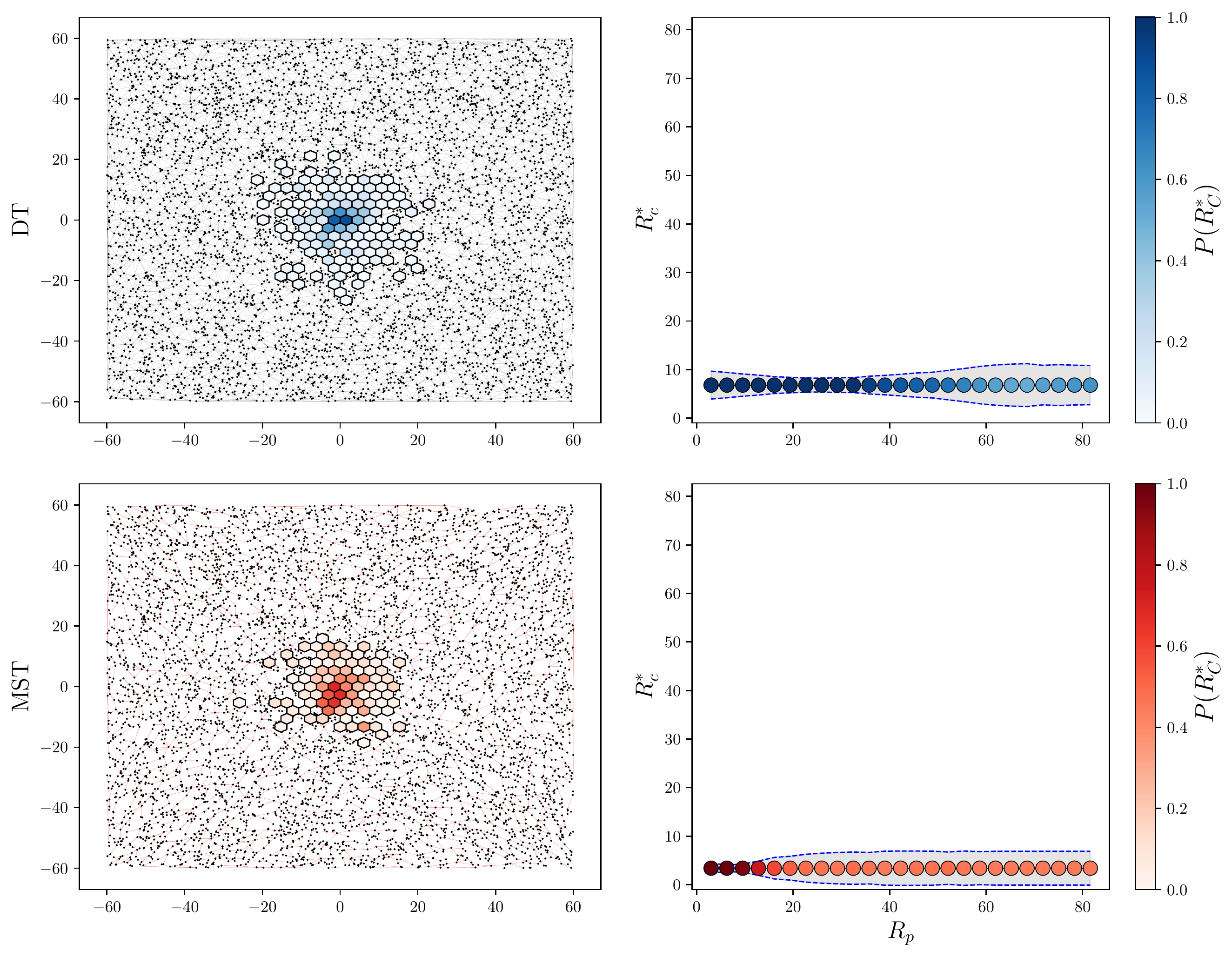}
  \caption{\label{fig:NullModels} Spatial behavior of congestion for a DT (top) and a MST (bottom) networks, built on a uniform distribution of $N=6000$ nodes in a square with side length $L=60$. Left column contains the graphic representations of the networks. Hexagonal bins provide information about the occurrence of congestion nodes in space, and their colors show the related frequency in proportion to the intensity. In the right column we present the dependence of the congestion radius $R_c$ on the patch radial size $R_p$, i.e., the radius of the circle centered in the middle of the network, that is used to define the considered subgraph within this radius. The congestion radius range is divided in bins, and each point is located at the statistical mode $R^*_c$ obtained with the distribution of $R_c$ after 100 realizations.  The color of the circular markers shows the probability of that value over the experimental $R_c$ distribution.  Shadow areas represent the variance of the different realizations of $R_c$ values with respect to the bin average value. Importantly, congestion mostly occurs in the center and no transitions arise, suggesting that these are a consequence of the intertwining of different graph structures.}
\end{figure}

We study first the behavior of congestion in the center-periphery intertwined framework introduced above. The quantitative assessment of congestion may be performed by recalling betweenness centrality. There exists a large set of studies about traffic \cite{Guimera2002,zhao2005onset,Echenique2005,yan2006efficient,liu2007method,dong2012enhancing,tan2014traffic,manfredi2018mobility} where the critical injection rate of vehicles $\gamma_c$, i.e., the maximum rate at which vehicles can enter the system without congesting it, can be evaluated as
\begin{equation}\label{eq:CongParam}
  \gamma_c = \frac{N-1}{B_n^{\ast}}\,.
\end{equation}
where $N$ is the nodes number of the street network and $B_n^{\ast}$ represents the maximum node betweenness.

The analysis of congestion in terms of betweenness permits to evaluate how congestion is affected by the underlying network structure. It is worth highlighting that the definition of betweenness embodies a routing protocol which implicitly assumes a traffic dynamic, relying on shortest-paths, and an Origin-Destination matrix which in our case is assumed to be uniform and all-to-all. Thus, combining these two ingredients the traffic model can be defined with precision. Previous works \cite{sole2018decongestion, Sole2016, sole2016model, sole2019effect} show the exact correspondence between the betweenness centrality as a measure to estimate node and edge traffic load, and the results obtained through the use of agent-based models (the actual traffic model).

Equation~\eqref{eq:CongParam} suggests that the maximum betweenness node marks the congestion onset. We are interested in its geographic location, that can be quantified by the distance to the center (radius) of the maximum betweenness node, hereafter called the congestion radius, $R_c$. In order to explore how the congestion location is affected by the intertwining of arterial and central roads, we calculate $R_c$ for subgraphs of the DT-MST synthetic networks, contained in concentric circles with radius $R_p$. Low values for this patch radius $R_p$, e.g., $R_p<R^{(DT)}$ refer to situations where the arterial periphery is suppressed, while large values, i.e., $R^{(DT)}<R_p \lesssim L$ describe an urban context where arterial periphery is relevant.

Even though there is not a direct mapping, there are at least two different urban processes that may be simulated by tuning the patch radius $R_p$: city sprawl, and changes in daily traffic patterns. In this latter case, the patch radius $R_p$ defines the city geographical domain where traffic dynamics is concentrated, and out of which it is supposed to be negligible. Large values for this quantity describe situations in which peripheral traffic flow is significant.

The dependence of the congestion radius $R_{c}$ on the patch radius $R_p$ is presented in Fig.~\ref{fig:NullModels} (right) for the DT and MST null models. In both situations, the congestion radius is constant and exhibits a very low value compared with the size of the system $L$. This means that congestion mainly occurs in the city center, regardless of the selected urban area. Such behavior is also reflected in the left panel, where hexagonal bins provide an information about the location of maximum betweenness nodes.

\begin{figure}[tb!]
  \includegraphics[width=0.95\columnwidth]{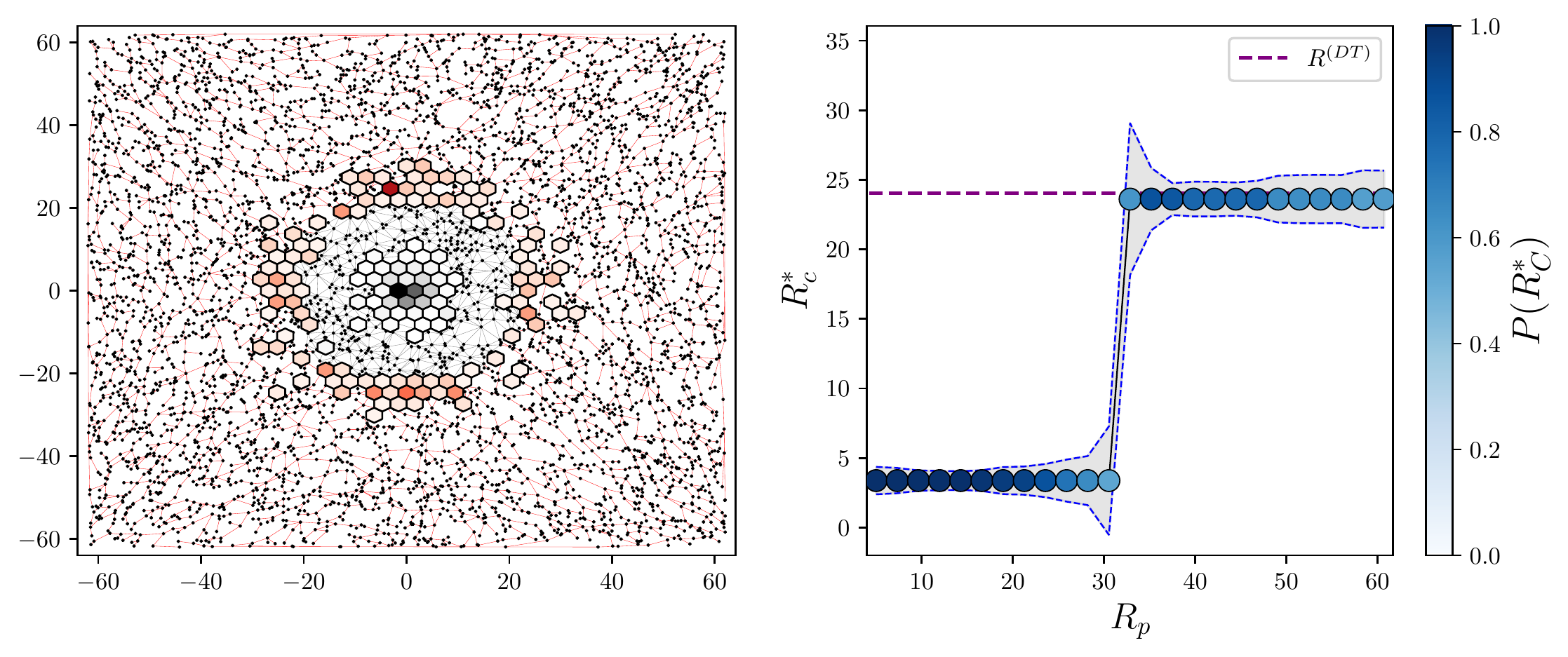}
  \caption{\label{fig:DT_MST_random_planar_model} Spatial behavior of congestion nodes for the random planar models introduced in Sec.~\ref{sec:DT_MST_model} with $N=6000$, $L=60$, and $R^{(DT)}=24$. In the left panel, the resulting network configuration, with MST and DT edges painted in red and black, respectively. Hexagonal bins provide information about the occurrence of congestion nodes in space, and their colors show the related frequency in proportion to the intensity. In the right panel,  we present the dependence of the congestion radius $R_c$ on the patch radial size $R_p$, i.e., the radius of the circle centered in the middle of the network, that is used to define the considered subgraph within this radius.  The congestion radius range is divided in bins, and each point is located at the statistical mode $R^*_c$ obtained with the distribution of $R_c$ after 100 realizations.  The color of the circular markers shows the probability of that value over the experimental $R_c$ distribution.  Shadow areas represent the variance of the different realizations of $R_c$ values with respect to the bin average value.}
\end{figure}

In Fig.~\ref{fig:DT_MST_random_planar_model} we repeat the same experiment but for the DT-MST random planar model introduced in Sec.~\ref{sec:DT_MST_model}. The relationship between congestion radius and patch one is not constant anymore, but depicts a step pattern configuration, namely an abrupt transition between two regimes associated to congestion at the DT center and at its connection with the MST. The former emerges for small patches size, while the latter arises at large ones, i.e., when the arterial roads contribute significantly to the network traffic. The comparison with Fig.~\ref{fig:NullModels} suggests that this unexpected behavior is due to the interplay of the composing structures, and in particular to the spatial discontinuity produced in the network.  Note that one could also implement the DT-MST entanglement by homogeneously overlapping the edges of the former an arboreal skeleton, without any spatial separation, and, as discussed in \cite{Lampo2021} (Fig.\ S10), no transition is detected. This is a crucial aspect that will be explored in the rest of the paper.

\section{Center-periphery entanglement with spatial overlap}
\label{sec:Spatial_Overlap}

The spatial discontinuity between the two topologies, resulting from their embedding on separated regions, is a fundamental ingredient to obtain the abrupt transition. So far we have considered a sharp separation between DT and MST. However, this is an extremely idealized situation, since the transition between a dense center to a sparse periphery usually occurs in a wider spatial range.

In this section we maintain the general separation between DT and MST but include the possibility of a spatial overlap in a finite interval. This is obtained by introducing a hybrid region with intermediate connectivity properties. The new region is ring-shaped, with small and large radius respectively given by $R^{(DT)}$ and $R^{(DT)}+\Delta R$, where $\Delta R>0$ denotes the width of the ring. In this region, we cover the MST skeleton with a density $\pi\in\left[0,1\right]$ of DT edges, sampled from a uniform distribution. The extreme bound $\pi=0$ refers to the situation in which no edge has been added, while $\pi=1$ corresponds to a completely saturated region.

\begin{figure}[tb!]
  \includegraphics[width=0.90\columnwidth]{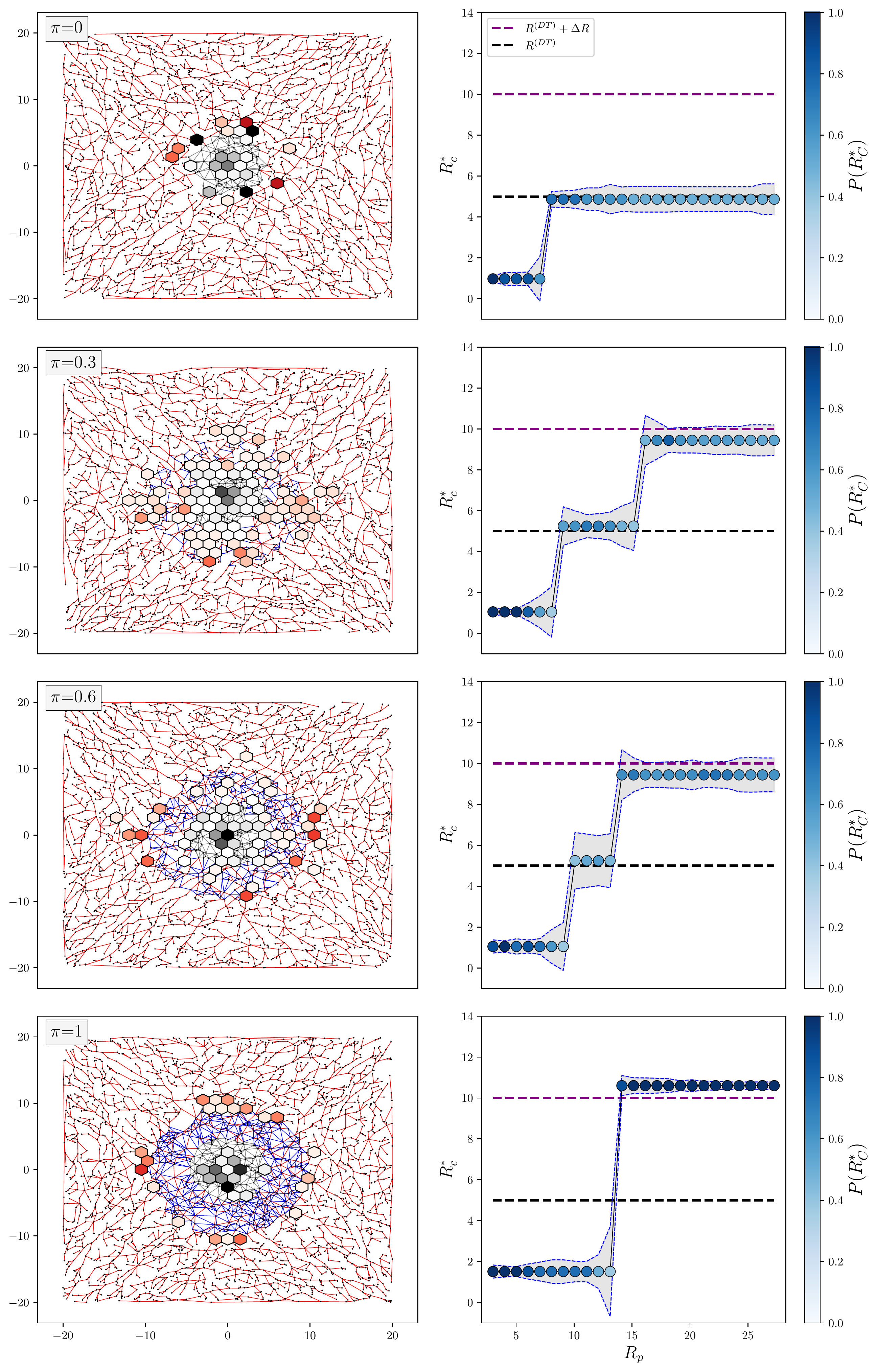}
  \caption{\label{fig:SpatialOverlap} Spatial behavior of the congestion nodes for the setup introduced in Sec.~\ref{sec:Spatial_Overlap}.  The figure is organized similarly to Figs.~\ref{fig:NullModels} and~\ref{fig:DT_MST_random_planar_model}: left panels show the network configurations, where blue edges refer to the $\pi$ density of further $DT$ links we add on top of the MST skeleton in the $[R,R+\Delta R]$ hybrid region. Still, hexagonal bins describe the occurrence of congestion in space. Right panels presents the dependence of congestion radius on the patch one. Here, circles are located at the statistical mode over an ensemble of 100~realizations of different random distributions of $N=3000$ nodes in a square of side equal to $L=20$, with $R^{(DT)}=5$, and $\Delta R=5$. The figure highlights the emergence of a further intermediate congestion regime, as a consequence of the hybrid region marked in blue.
}
\end{figure}

This class of networks is shown in the left column of Fig.~\ref{fig:SpatialOverlap}, for different values of the parameter $\pi$. The corresponding relationship between the congestion radius and the patch one is presented in the right column.  The cases with extreme values $\pi=0$ and $\pi=1$ fall into the category treated in the previous section, with DT radius equal to $R^{(DT)}$ and $R^{(DT)}+\Delta R$, respectively. These exhibit one abrupt transition between two regimes related to congestion in the center and in its frontier with the periphery.

Interesting phenomenology arises for intermediate values of $\pi$.  In these cases, the hybrid region is not saturated and differs from both the DT and MST. This induces a further structural discontinuity in the network, which affects the spatial behavior of congestion. Particularly, a second abrupt transition stands out: as $R_p$ increases, congestion points abruptly shift from the center to the DT frontier ($R_c\approx R^{(DT)}$), and then to that of the hybrid ring ($R_c\approx R^{(DT)}+\Delta R$). The spatial overlap, thus, induces a further possible congestion location.

This experiment sheds light on the mechanism producing the abrupt transition. The structural discontinuity required to obtain such a phenomenon consists of a sharp change of the spatial density of edges, i.e., number of edges over the unit area. Then, one may suppose that such quantity permits to manipulate the abruptness of the congestion transition; this is the aim of Sec.~\ref{sec:Edge_Dist}. Additionally, the spatial edge density can also be modified by acting on the nodes distribution, rather than network connectivity; this will be explored in Sec.~\ref{sec:NodeDens}.

The intermediate congestion regime emerged in Fig.\ \ref{fig:SpatialOverlap} portrays a similar pattern to that detected in \cite{Lampo2021}, although in a different framework: GT-model. This consists of square grid, representing the city center, connected to four regular trees, one at each side, representing the urban periphery. 
In this context, three congestion regimes arise, localized respectively in the grid center, grid side and the tree root. 
This situation is equivalent to that discussed in the current section. the GT-model, indeed, defines three regions with different edges density: the areas where are embedded the grid, the trees and the grid-tree connections. The discontinuity between these regions lead to a multiple congestion pattern. 
 
\section{Center-periphery entanglement with non-uniform edges distribution}
\label{sec:Edge_Dist}

In the previous section we recognized the relationship between the observed phenomenology and the connectivity properties of a road network. Particularly, abrupt transitions follow from sharp discontinuities in the edge density. This suggests that bottlenecks location may be controlled by properly tuning such quantity. This is the purpose of the current section.

So far, the spatial discontinuity in the edge density emerged spontaneously as a consequence of the intertwining of different topologies. However, within a given region the connectivity has been considered constant in space. Particularly, the hybrid region introduced in Sec.~\ref{sec:Spatial_Overlap} was constructed by covering the MST skeleton with DT edges sampled from a uniform spatial distribution. This constitutes a strong idealization because road density is in general not uniform over the city. Specifically, the transition from a dense center to a sparse arterial periphery depicts a non-trivial profile of the street distribution, interpolating between that of the DT and the MST one.

We propose an extension of the setup in Sec.~\ref{sec:Spatial_Overlap}, where the DT edges in the hybrid region are sampled from a non-uniform distribution. We assume that an edge with center at radius $r$ has a probability to be added given by
\begin{equation}\label{eq:DensProfile}
  P(r)=\alpha\exp(-kr)+\beta,\quad r\in\left[R^{(DT)},R^{(DT)}+\Delta R\right],
\end{equation}
where $k\neq0$ is a real constant ruling the damping in space of the exponential. The $\alpha$ and $\beta$ coefficients are fixed by the boundary conditions
\begin{equation}
  P(R^{(DT)})=1,\quad P(R^{(DT)}+\Delta R)=0\,,
\end{equation}
which lead to
\begin{equation}\label{eq:Coeff}
  \alpha=\frac{1}{e^{-kR^{(DT)}}-e^{-k(R^{(DT)}+\Delta R)}},\quad\beta=\frac{1}{2}\left[1-\frac{e^{-kR^{(DT)}}+e^{-k(R^{(DT)}+\Delta R)}}{e^{-kR^{(DT)}}-e^{-k(R^{(DT)}+\Delta R)}}\right].
\end{equation}
Additionally, for generalization purposes and consistency with the method in Sec.~\ref{sec:Spatial_Overlap}, we ensure that the final network includes a fraction $\pi$ of all the candidate edges.

\begin{figure}[tb!]
  \includegraphics[width=0.95\columnwidth]{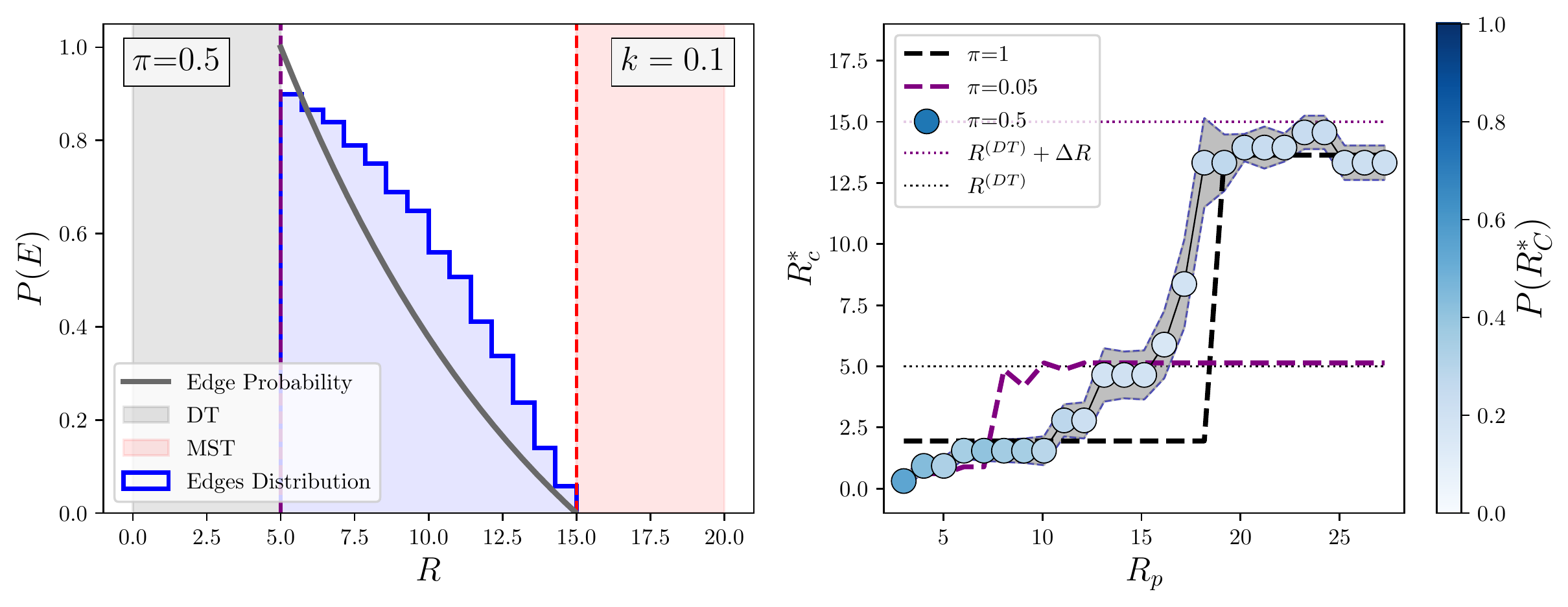}
  \caption{\label{fig:EdgeDensity1} Congestion radius pattern for the setup introduced in Sec.~\ref{sec:Edge_Dist}, with $N=3000$, $L=20$, $R^{(DT)}=5$, and $\Delta=10$. Hybrid region is generated by overlapping a fraction $\pi=0.5$ of DT edges sampled from the distribution in Eq.~\eqref{eq:DensProfile}, with $k=0.1$. In the left panel, the blue histogram describes the final distribution of these edges. The right panel reports the congestion radius $R_c$ as a function of the parch radius $R_p$. The black and purple dashed line report the result related to $\pi=1$ and $\pi=0.05$ respectively, where all the edges and approximately no one have been added, and match the situation described in Sec.~\ref{sec:Spatial_Overlap}. The dotted line refers to the edge distribution shown in the right panel, where circles are located at the statistical mode over an ensemble of 100~realizations. In this case transition exhibits a smooth, rather than abrupt, profile as a consequence of the non-tight discontinuity in the edge density.}
\end{figure}

This procedure introduces a bias in the construction of the hybrid region, whose edge connectivity interpolates now between that of the DT and the MST. In Fig.~\ref{fig:EdgeDensity1} (left) the grey solid line refers to Eq.~\eqref{eq:DensProfile} for a given choice of the parameter $k$. The blue histogram portrays the final distribution of a fraction $\pi=0.5$ of the DT edges: their density approximately reproduces the profile in Eq.~\eqref{eq:DensProfile} and is not constant in space but decreases exponentially as a function of the radius. Of course, when $\pi=1$ the hybrid region saturates and gets indistinguishable from the DT core, which extends until $r=R^{(DT)}+\Delta R$. In this case the system reduces to the DT-MST model in Sec.~\ref{sec:DT_MST_model}. In the right panel we present the dependence of the congestion radius $R_c$ as a function of the patch size $R_p$. Black (purple) dashed lines refer to $\pi=1$ ($\pi=0.05$), corresponding to a hybrid region in which all the edges (approximately no one) have been added, and reproduces the abrupt transition pattern discussed in Sec.~\ref{sec:DT_MST_model}. A new behavior arises at $\pi=0.5$: the transition between different congestion regimes is not abrupt anymore. Congestion radius, approximately equal to zero at small parches, smoothly grows until $r=R^{(DT)}+\Delta R$, as a consequence of the continuous edge density degrowth shown in the left panel.

\begin{figure}[tb!]
  \includegraphics[width=0.95\columnwidth]{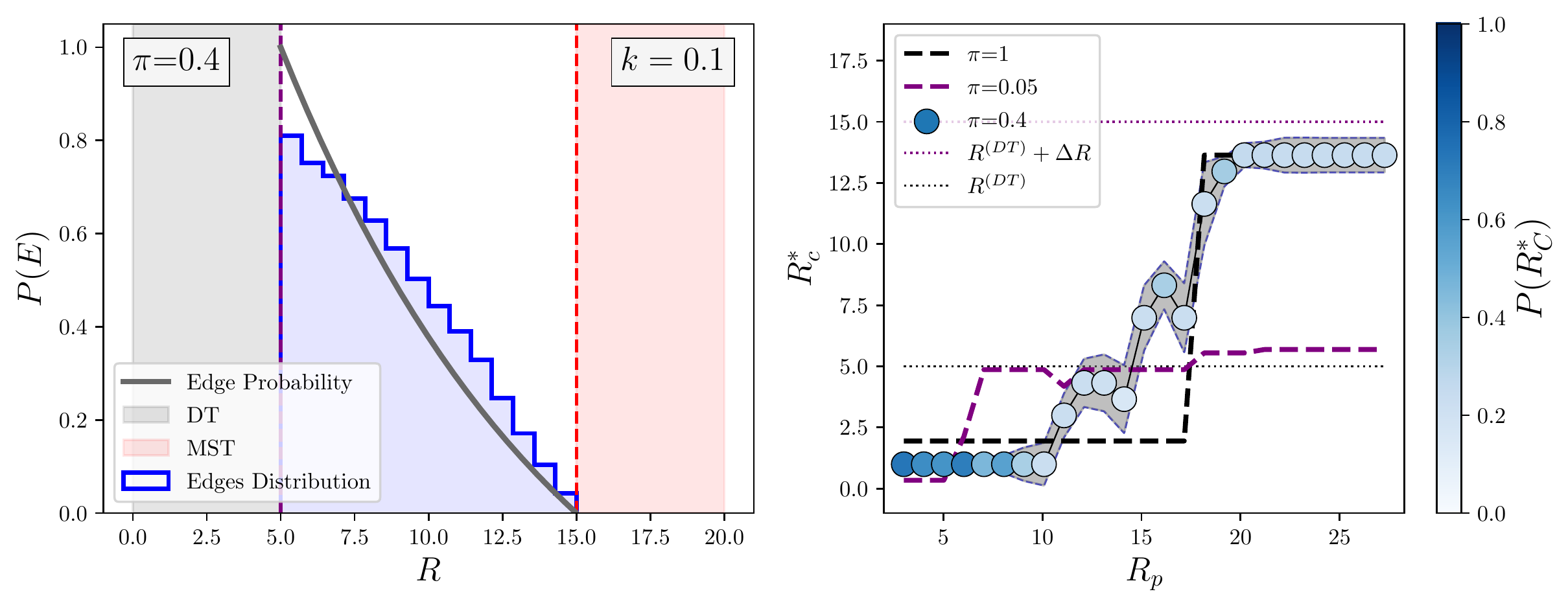}
  \caption{\label{fig:EdgeDensity2} Congestion radius pattern for the setup introduced in Sec.~\ref{sec:Edge_Dist}, with $N=6000$, $L=20$, $R^{(DT)}=5$, and $\Delta=10$. Hybrid region is generated by overlapping a fraction $\pi=0.4$ of DT edges sampled from the distribution in Eq.~\eqref{eq:DensProfile}, with $k=0.1$. In the left panel, the blue histogram describes the final distribution of these edges. The right panel reports the congestion radius $R_c$ as a function of the parch radius $R_p$. The black and purple dashed line report the result related to $\pi=1$ and $\pi=0.05$ respectively, where all the edges and approximately no one have been added, and match the situation described in Sec.~\ref{sec:Spatial_Overlap}. The dotted line refers to the edge distribution shown in the right panel, where circles are located at the statistical mode over an ensemble of 100~realizations. In this case transition exhibits a smooth, rather than abrupt, profile as a consequence of the non-tight discontinuity in the edge density.}
\end{figure}

A similar behavior is detected in Fig.~\ref{fig:EdgeDensity2}, for a different choice of parameters. Both make use of $k=0.1$, and this deserves to be commented. High values for $|k|$ make the function in Eq.~\eqref{eq:DensProfile} elbow-shaped, leading to a quasi-sharp second jump in the edge density, which would reproduce the situation analyzed in Sec.~\ref{sec:Spatial_Overlap}. In general, the density profile in Eq.~\eqref{eq:DensProfile}, with the coefficients in Eq.~\eqref{eq:Coeff}, approaches a constant function with value~1 for $k\rightarrow-\infty$, and with value~0 for $k\rightarrow+\infty$. These two cases fall into the situation described in Sec.~\ref{sec:Spatial_Overlap}, with a DT radius equal to $R^{(DT)}+\Delta R$ and $R^{(DT)}$, respectively.

\section{Center-periphery entanglement with non-uniform nodes distribution}
\label{sec:NodeDens}

The analysis of the previous sections suggests that the existence of multiple congestion regimes follows from different connectivity regions in the road networks. In the most simple case of Sec.~\ref{sec:DT_MST_model}, this has been implemented by entangling a DT core and a MST periphery, yielding a discontinuity in the edge density and so to different degrees of paths redundancy. We argue that there exist alternative implementations to reproduce this situation.

\begin{figure}[tb!]
  \includegraphics[width=0.97\columnwidth]{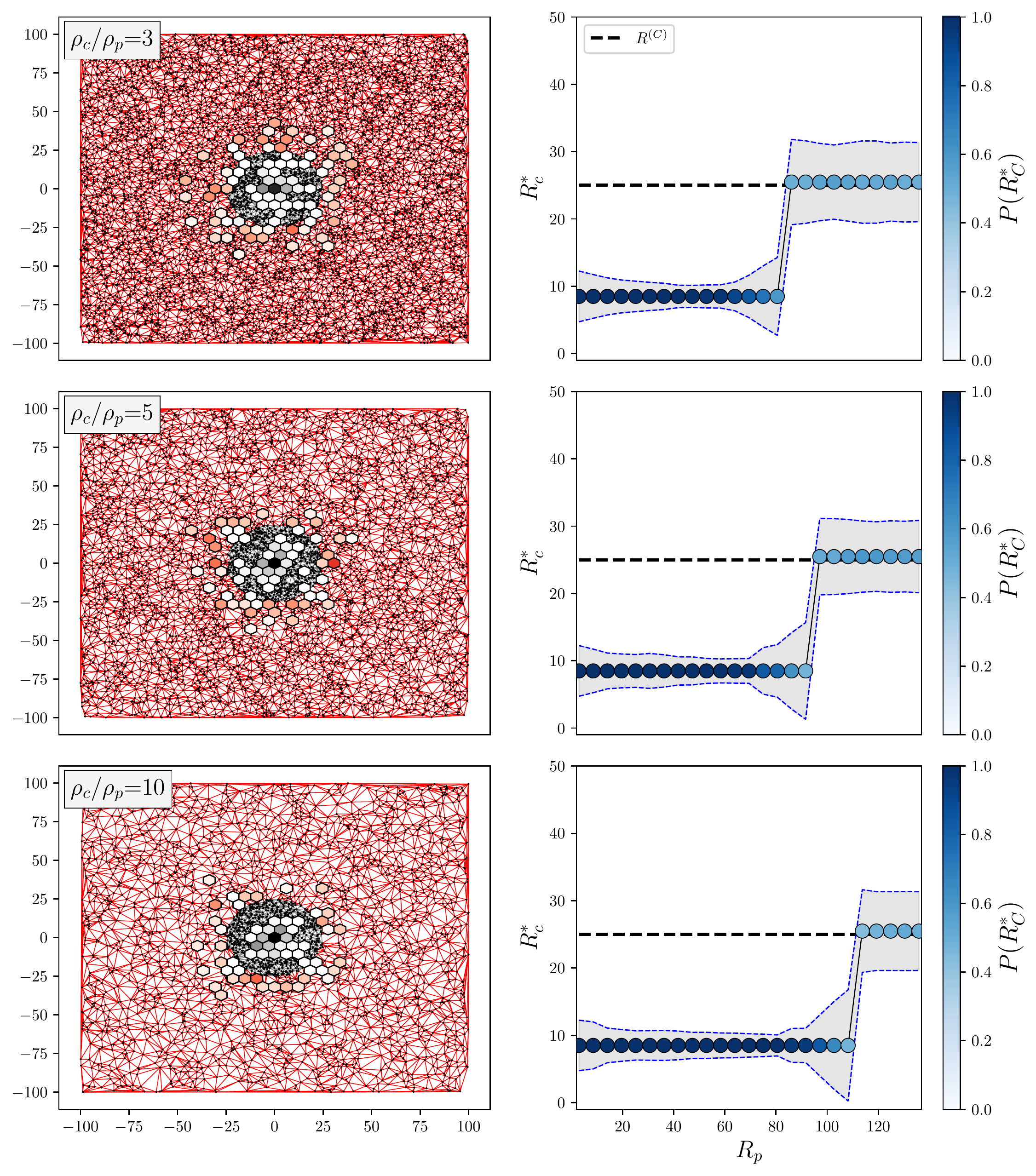}
  \caption{\label{fig:NodeDens} Spatial behavior of congestions for the network model introduced in Sec.~\ref{sec:NodeDens}. Left column shows the network configuration, where center and peripheral edges are painted in black and red, respectively. Here, hexagonal bins provide an information about the location of maximum betwenness nodes. Right column presents the dependence of the congestion radius on the patch one. Here, circles are located at the statistical mode over an ensemble of 100~realizations of different random distributions of nodes in a square of side equal to $L=100$, with $R^{(c)}=5$, $\rho_c=\frac{N}{\pi (R^{(C)})^2}$, ($N=10000$). Rows refer to different density ratio. The figure depicts a clear transition pattern as that detected in Fig.~\ref{fig:DT_MST_random_planar_model}}.
\end{figure}

So far, the network model has been constructed over an uniform distribution of nodes, i.e., a constant nodes density. This leads to considering different topologies in order to produce the wished change in the spatial density of edges. One could also look into the inverse situation, in which center and periphery result from different nodes densities, high in the center and low in the periphery, but exhibiting the same connectivity properties (i.e., the same average degree).

The network we consider in this section has a DT structure, and a different nodes density between center and periphery. The former, defined as a circle with radius $R^{(C)}$, has density $\rho_c$, while that of the latter is $\rho_p<\rho_c$. In the limit in which $\rho_c=\rho_p$ the network reduces to a uniform DT as that presented in Fig.~\ref{fig:NullModels}. We assume that both $\rho_c$ and $\rho_p$ are constant, i.e., they are uniform within their own regions.

In Fig.~\ref{fig:NodeDens} we present the spatial behavior of congestion for different center-periphery density ratios $\rho_c/\rho_p$. It is possible to note a clear abrupt transition between a regime describing congestion in the center, and its connection with the periphery. We remark once more that the periphery is not modeled as a MST, but the whole network results from a DT over the nodes distribution. This confirms the possibility to produce an abrupt displacement of traffic bottlenecks as a consequence of a discontinuity in the nodes density, without any change in the topology.

An important aspect of the behavior in Fig.~\ref{fig:NodeDens} concerns the pattern at different density ratios $\rho_c/\rho_p$. The minimum patch radius value allowing to observe congestion in the center-periphery connection, grows as the density ratio increases. This is due to the fact that, in addition to a structural discontinuity, the transition requires a large peripheral flux (i.e., a sufficiently large number of nodes) in order to emerge. For a given choice of the system parameters, such flux reduces as the density ratio decreases. We expect, indeed, that when the peripheral number of nodes is low enough, $\rho_c/\rho_p\rightarrow\infty$, no transition arises, since we approach a network compound by a DT core with a few surrounding points. Therefore, the experiment discussed in this section is constrained to a certain range values of $\rho_c/\rho_p$: large enough to produce a structural discontinuity, low enough to ensure a relevant peripheral flux.

\section{Conclusions and Perspectives}
\label{sec:Conclusions}

In \cite{Lampo2021} we showed that traffic bottlenecks shift away from the city center as larger concentric areas are included in the urban spatial domain, giving rise to abrupt congestion transitions. The current work sheds lights on the main mechanisms which permit to control the phenomena, and provides useful insights in urban planning and the management of traffic.

The analysis in \cite{Lampo2021} points to the structural discontinuity between central and arterial roads as the fundamental ingredient behind the abrupt transition of congestion points. The present paper clarifies the formal meaning of such structural discontinuity and relates it to the spatial density of edges. We show that, acting on this quantity, it is possible to modify the transition pattern, and so to control the geographic occurrence of congestion. Briefly, our results may be summarized as follows. First, the introduction of additional jumps in the edge density is associated to the emergence of new abrupt transitions, (Sec.~\ref{sec:Spatial_Overlap}). Second, these transitions get smoother if the space-density of DT edges on the MST skeleton exhibits a proper continuous profile (Sec.~\ref{sec:Edge_Dist}). Third, abrupt transitions may be obtained by means of discontinuities in the nodes density, without any change in the network topology.

These results extend the applicability of the DT-MST model and allows to better interpret the empirical validation presented in \cite{Lampo2021} (Sec.~VII). There, the congestion radius behavior as a function of the parch size was studied for street networks of 97 cities worldwide. Overall results showed that 52~cities present clear, detectable at naked eye, regimes with abrupt transitions between them. The existence of multiple transitions, rather than a single one, may be understood as the consequence of various edge density discontinuity points in the road networks. Still, 45~cities present detectable regimes with smoother transitions which can be now understood as the results of smooth change in the edge density.

The possibility to control the spatial distribution of the congestion points, and in particular to remove them from the center, constitutes an interesting tool for urban planning. The displacement of the traffic bottlenecks towards peripheral areas is a positive effect to reduce pressure on the center, which is usually not prepared to support high vehicles loads. Within this context, excessively abrupt shift of bottlenecks indicates areas where the structural transition is excessively sharp, and this must have implications, yet to be studied, with respect to the  efficiency (in whatever terms) of the transportation system.

The setup we have developed constitutes a complete benchmark to address several problems related with the spatial localization of bottlenecks. First, the role of edge weights deserves to be better investigated. In this work we consider these as the Euclidean distances between the endpoints, but it would be interesting to wonder if the congestion location may be manipulated by means of the weights distribution. Secondly, concerning the reproduction of the aforementioned empirical results, much effort is needed to understand which actually is the effective origin of transition beyond edge density: discontinuity in nodes distribution, or in connectivity? Our framework permits to construct synthetic networks with nodes and edges properties sampled from the corresponding empirical distributions, thus constituting a simple playground to scrutinize the relationships between structure and function in urban road networks.

\section*{Declarations}

\begin{backmatter}

\section*{Availability of data and materials}
The codes used during the current study are available from the corresponding author on reasonable request.

\section*{Competing interests}
The authors declare that they have no competing interests.

\section*{Funding}
A.L., J.B-H.\ and A.S-R.\ acknowledge the support of the Spanish MICINN project (grant PGC2018-096999-A-I00). S.G.\ acknowledges financial support from Spanish MINECO (grant PGC2018-094754-B-C21), Generalitat de Catalunya (grant 2017SGR-896), and Universitat Rovira i Virgili (grant 2019PFR-URV-B2-41).

\section*{Authors' contributions}
All the authors contributed equally.

\section*{Acknowledgements}
Not applicable

\section*{Abbreviations}
\begin{itemize}
\item DT: Delaunay Triangulation
\item MST: Minimum Spanning Tree
\end{itemize}


\bibliographystyle{bmc_mathphys}
\bibliography{references}

\end{backmatter}

\end{document}